\documentclass[a4paper,11pt]{article}
\usepackage{pos}

\title{Analytic results for massive $2\to2$ processes}

\author*[a]{Matthias Steinhauser}

\affiliation[a]{Institut f{\"u}r Theoretische Teilchenphysik,
    Karlsruhe Institute of Technology (KIT),\\
  Wolfgang-Gaede Stra\ss{}e 1, 76128 Karlsruhe, Germany}

\emailAdd{matthias.steinhauser@kit.edu}

\abstract{We discuss recent (semi) analytic results for $2\to 2$ processes
  with massive internal and external particles in various regions of phase
  space. In the physical applications we restrict ourselves to $gg\to HH$.}

\FullConference{16th International Symposium on Radiative Corrections: Applications of Quantum Field Theory to Phenomenology (RADCOR2023)\\
28th May - 2nd June, 2023\\
Crieff, Scotland, UK\\}

\begin{document}
\maketitle


\section{Introduction}

At this conference many results for multi-loop calculations have been
reported. Often the internal particles are all massless which often also
holds for most of the external lines~\cite{proceedings_RADCOR_2023}.
However, there are also processes where it is important to keep the
mass of the internal particles. A prime example is Higgs boson pair
production in gluon fusion but also processes like $gg\to ZZ$, $gg\to
ZH$ or Higgs plus jet production. Furthermore, once electroweak
corrections are considered the gauge and Higgs boson masses occur in
internal lines.

In general analytic results for massive $2\to 2$ processes are rare
and, if available (see, e.g., Refs.~\cite{Bonciani:2019jyb,Becchetti:2023wev}),
they have a complicated analytic structure
which makes them often difficult to handle and the numerical evaluation
is not straightforward.

On the other hand, there are purely numerical approaches, which are
often computationally expensive. Furthermore, numerical results are
significantly less flexible, e.g., in connection to changes of
renormalization schemes.

In this contribution we discuss analytic approximations. Individual
results are valid in certain limits. However, their combination can
cover the whole phase space. The approximations which are
discussed in the following consist either of analytic expansions, which
are composed of simple functions, or of power-log expansions with
precise numerical coefficients. In either case, a fast numerical
evaluation is guaranteed.  Needless to say, analytic approximations
allow for a flexible use, in particular in connection to
renormalization scheme changes.

In this proceedings contribution we discuss several recent results for
massive $2\to 2$ processes at two and three loops. We concentrate on
$gg\to HH$. The techniques can also be applied to other processes such
as gauge and Higgs boson production in gluon fusion or Higgs plus jet
production. We will not discuss the large-$m_t$ expansion which for
$gg\to HH$ is available up to
NNLO~\cite{Grigo:2015dia,Davies:2019djw,Davies:2021kex}. Recently
also the full electroweak corrections have been computed in this
limit~\cite{Davies:2023npk}.  Exact NLO QCD results for $gg\to HH$ are
available from
Refs.~\cite{Borowka:2016ehy,Borowka:2016ypz,Baglio:2018lrj}.


\section{High-energy limit}

In Refs.~\cite{Davies:2018ood,Davies:2018qvx} analytic results for the
NLO QCD corrections for $gg\to HH$ have been obtained in the limit
$s,t,u \gg m_t^2 \gg m_H^2$. The second inequality sign leads to a
simple Taylor expansion, which can be performed at the integrand
level. This effectively eliminates the scale $m_H$. The remaining
integrals depend on $s,t$ and $m_t^2$ (with $u=-s-t$).  An expansion
for small top quark mass at the amplitude level is tedious. It is more
convenient to perform a reduction and implement $s,t\gg m_t^2$ at the
level of the
two-loop master integrals.

The results of Refs.~\cite{Davies:2018ood,Davies:2018qvx} have been
used in Ref.~\cite{Davies:2019dfy} to combine the high-energy
expansion with the exact numerical calculations of
Refs.~\cite{Borowka:2016ehy,Borowka:2016ypz} such that the latter has
to be used only in a restricted region of phase space. This 
significantly reduces the required CPU time.  For this analysis an
expansion up to $m_t^{32}$ was available.

\begin{figure}[t]
    \includegraphics[width=0.485\textwidth]{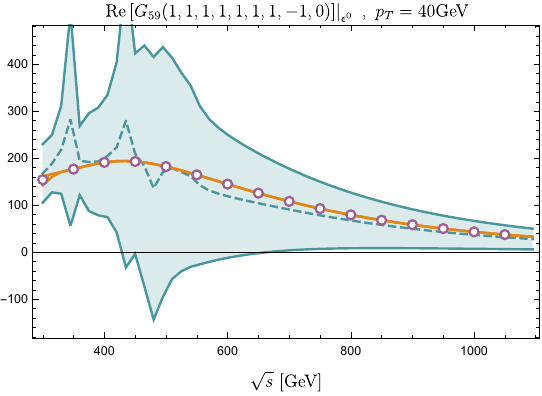}
    \includegraphics[width=0.485\textwidth]{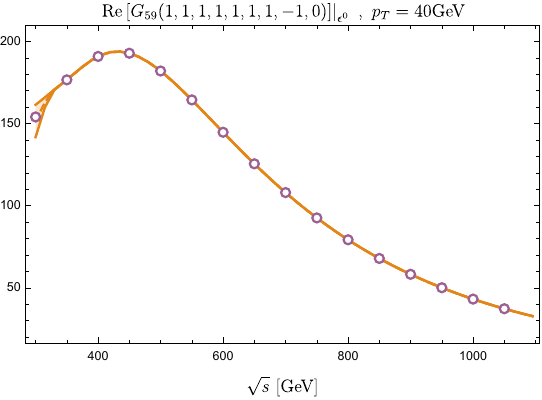}
    \\
    \centering
    \includegraphics[height=0.04\textwidth]{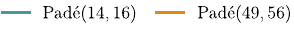}
    \includegraphics[height=0.04\textwidth]{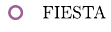}
    \\
    \flushleft
    \vspace{-5mm}\hspace{0.24\textwidth}(a)\hspace{0.47\textwidth}(b)
    \caption{Comparison of Pad\'{e}-based approximations constructed
      from different expansion depths ($N_{\rm low},N_{\rm high}$)
      with numerical results obtained using {\tt FIESTA}, for a
      non-trivial non-planar master integral, see
      Ref.~\cite{Davies:2023vmj} for details.
       \label{fig::non-pl_MI}
     }
\end{figure}

In Ref.~\cite{Davies:2022ram} the high-energy expansion has been
refined and deep expansions in $m_t^2/\{s,t,u\}$ could be obtained.
This leads to a significant qualitative improvement, in particular in
combination with the use of Pad\'e approximation which is applied to
the $m_t$ expansion. Using more then 100 expansion terms in $m_t$ we
can construct a large number of different Pad\'e approximants
together with an estimate of the uncertainty~\cite{Davies:2020lpf}.
The latter can be validated by comparing Pad\'e results of selected
scalar integrals to numerical results obtained with {\tt
  FIESTA}~\cite{Smirnov:2021rhf} or {\tt
  pySecDec}~\cite{Borowka:2018goh}. An example is shown in
Fig.~\ref{fig::non-pl_MI}.  The left panel shows in green high-energy
results based on 32 expansion terms and in orange results where 112
expansion terms are incorporated.  In both cases the central values and the
uncertainties are shown. Numerical results obtained with {\tt FIESTA}
are shown as circles. The magnification on the right panel shows the
impressive accuracy of the Pad\'e method, even close to the two top
quark threshold. A deep high-energy expansion accompanied with Pad\'e
approximation can thus be viewed as a precision tool for massive
$2\to 2$ Feynman integrals.

\begin{figure}[t]
  \centering
  \begin{tabular}{cc}
    \includegraphics[width=0.45\textwidth]{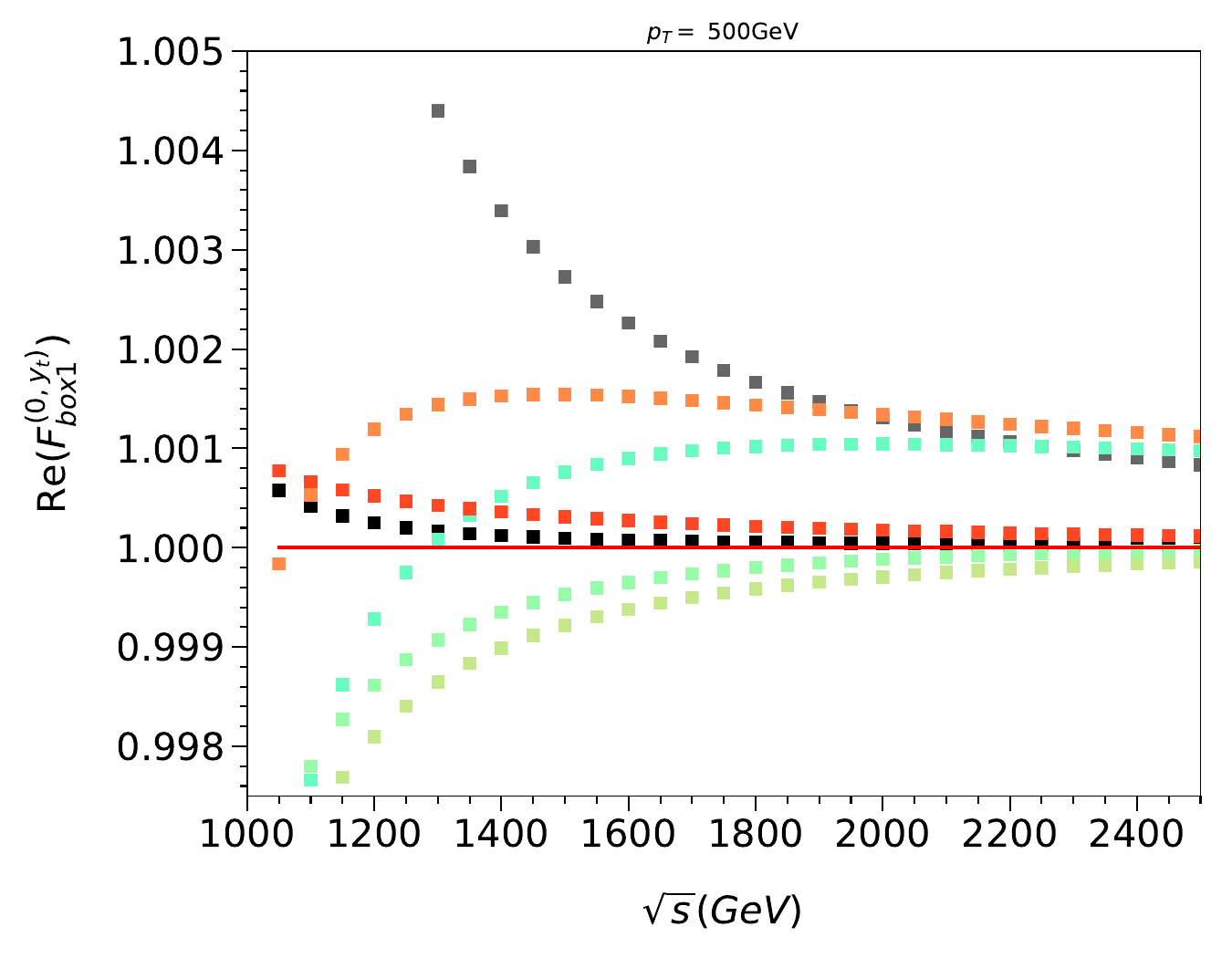}
    &
    \includegraphics[width=0.45\textwidth]{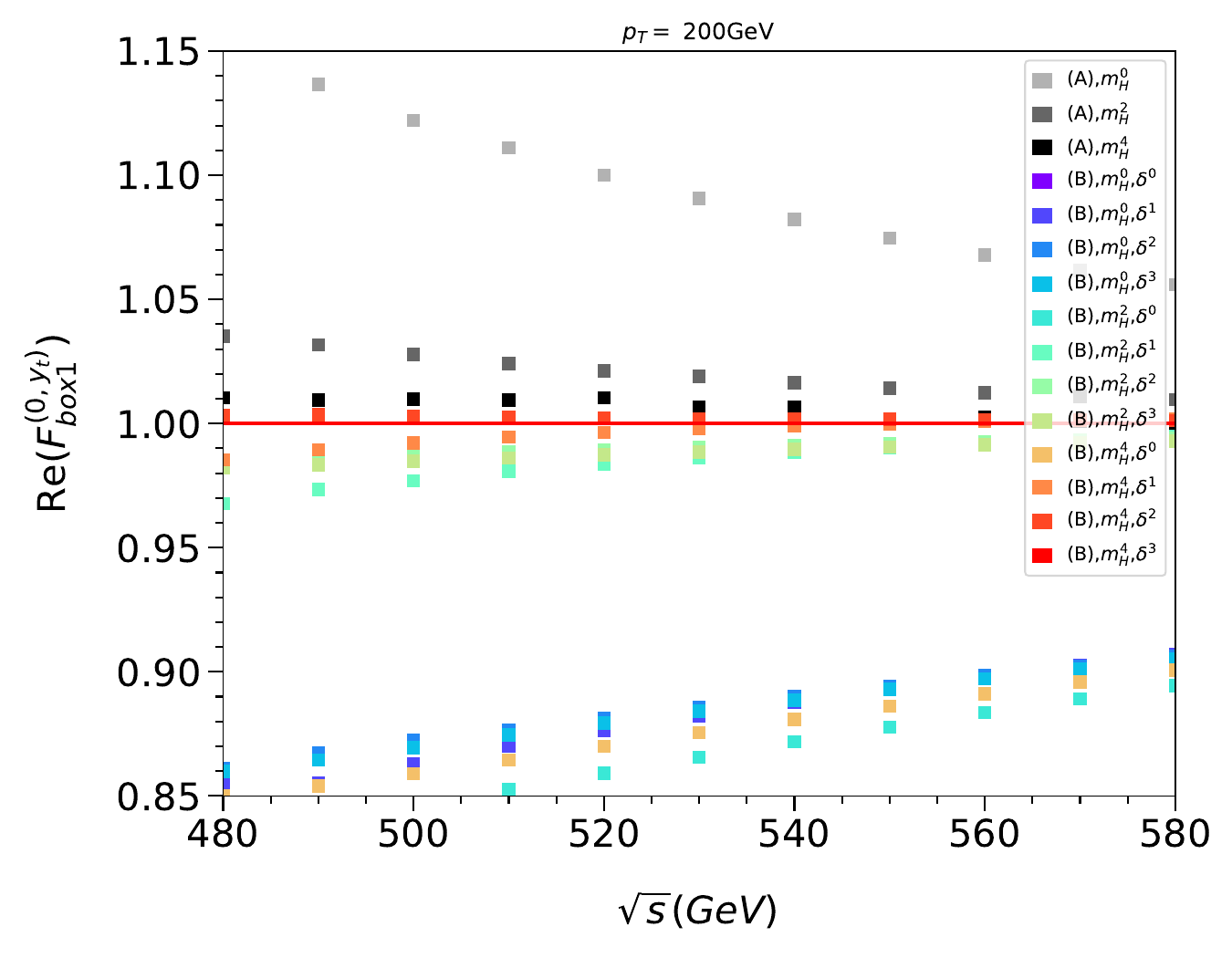}
      \\
    (a) & (b)
  \end{tabular}
  \caption{\label{fig::F12_ratio}Real part of $F_{\rm box1}$ for different
    values of $p_T$ and various expansion terms in
    $m_H^{\rm ext}$ and $\delta= 1 - m_H^{\rm int}/m_t$.}
\end{figure}

In Ref.~\cite{Davies:2022ram} a first step towards the electroweak
corrections has been taken and the high-energy expansion has been
applied to the two-loop box diagrams where a Higgs boson is exchanged
between the top quarks.
This introduces an
additional scale $m_H^{\rm int}$, the mass of the internal Higgs
boson. Exact analytic calculations are again most likely not possible
or at least quite involved.  On the other hand, one may
consider in addition to the high-energy limit either of the two cases:
\begin{itemize}
\item[(A)] $m_t^2 \gg (m_H^{\rm int})^2 = (m_H^{\rm ext})^2$,
\item[(B)] $m_t^2 \approx (m_H^{\rm int})^2 \gg (m_H^{\rm ext})^2$.
\end{itemize}
Here ``$m_t^2 \approx (m_H^{\rm int})^2$'' means a Taylor expansion in
the mass difference. In (A) ``$\gg$'' requires the application of an
asymptotic expansion in the large mass limit, which involves
non-trivial subdiagrams. On the other hand, in (B) ``$\gg$'' leads to
a Taylor expansion as above.  It has been shown in
Ref.~\cite{Davies:2022ram} that both cases lead to good results as can
be seen from Figs.~\ref{fig::F12_ratio}(a) and (b) where results for
the real part of $F_{\rm box1}$ are shown for $p_T=500$~GeV and
$p_T=200$~GeV, respectively. The colours correspond to approach~(B)
and the results from approach~(A) are shown in gray and black.  One
observes a nice convergence when including higher order expansion
terms and, furthermore, the results for the two approaches agree well
at or even below the percent level.

\section{$t\to0$ expansion}

In this section we still consider two-loop corrections to $gg\to
HH$. However instead of high-energy expansions we discuss expansions
around the forward scattering limit.  This idea has been applied for
the first time to $gg\to HH$ in Ref.~\cite{Bonciani:2018omm} under the name ``$p_T$
expansion''.  Later the method has been refined in
Refs.~\cite{Degrassi:2022mro,Bellafronte:2022jmo,Vitti:2022kot}.

In our approach~\cite{Davies:2023vmj} we Taylor-expand in the
Mandelstam variable $t$ and in the final-state masses independently,
whereas the $p_T$ expansion provides a combination of both
expansions. The two prescriptions are equivalent in the sense that
once the final result is expressed in terms of the same variables and
potential higher order terms are discarded one obtains identical
expressions.

\begin{figure}[t]
    \includegraphics[width=\textwidth]{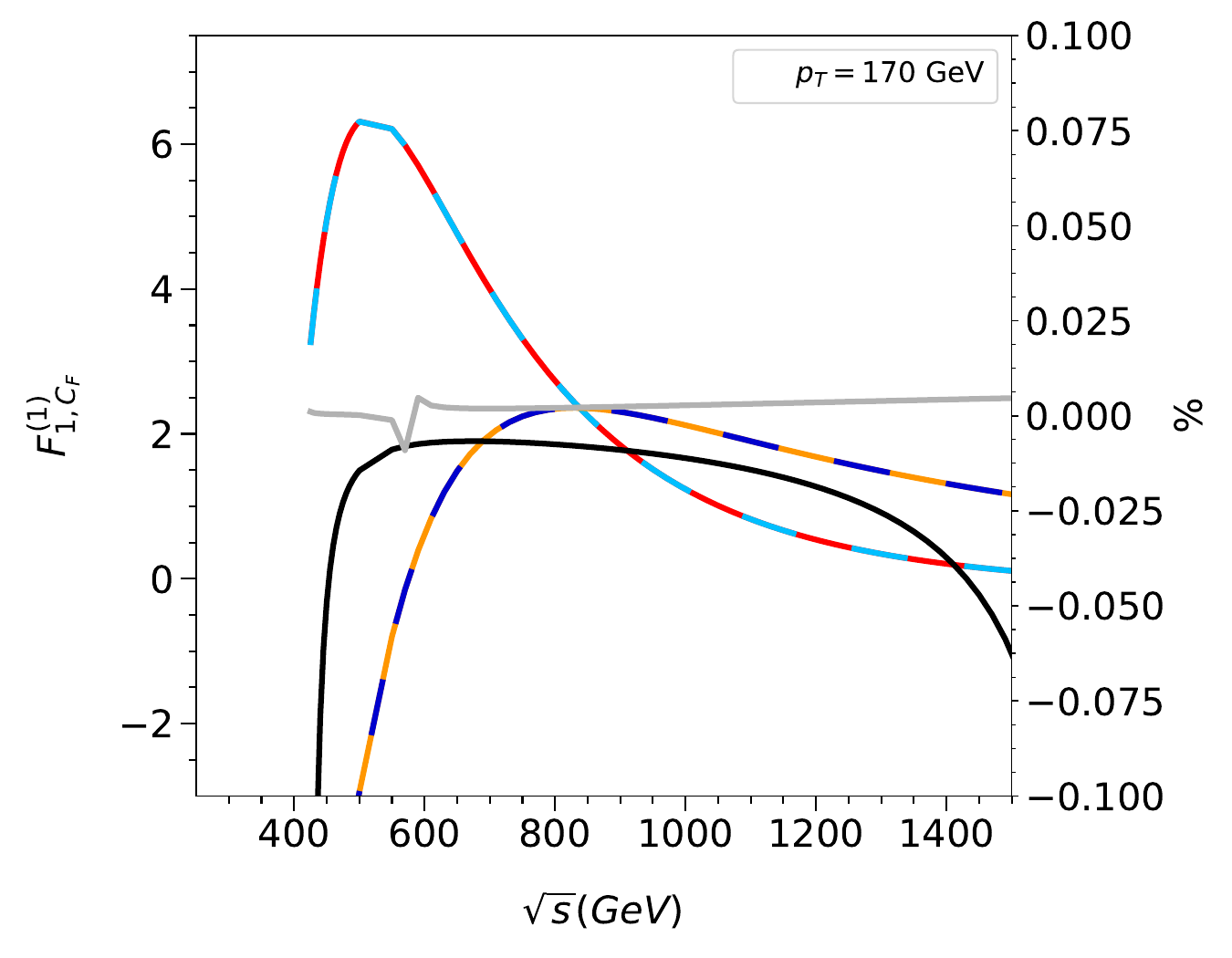} 
    \caption{\label{fig::F12l_cR}Real (red and light blue) and imaginary
      (orange and dark blue) parts of the $C_F$ contribution to the two-loop
      form factor $F_{\rm box1}^{(1)}$ (see Ref.~\cite{Davies:2018qvx} for a
      precise definition) as a function of $\sqrt{s}$ for $p_T=170$~GeV. The
      red and orange curves correspond to the expansion for $t\to 0$ and the
      light and dark blue curves to the high-energy expansion.  The relative
      differences (see scale on the right side) are shown as black (real
      part) and gray (imaginary part) curves.}
\end{figure}

In Ref.~\cite{Davies:2023vmj} we have shown that the expansion in $t$
converges very quickly for $p_T\lesssim 200$~GeV. Instead of order 100
terms, as for the high-energy expansion, only a few (we have computed
six terms) are sufficient. An expansion up to order $m_H^4$ is
sufficient to obtain results which show perfect agreement with the
exact expressions and deviate at most at the percent level for small
values of $\sqrt{s}$.  Furthermore, in Ref.~\cite{Davies:2023vmj} we
have shown that the combination of the high-energy and $t\to0$
expansion covers the whole phase-space and thus no purely numerical
approach is necessary (see also Ref.~\cite{Bellafronte:2022jmo} where
a similar approach has been proposed, though with less input from the
high-energy and around the forward limit).

In Fig.~\ref{fig::F12l_cR} we show the results for the $C_F$ part of
one of the form factors for $p_T=170$~GeV as a function of
$\sqrt{s}$.  For the small-$t$ expansion (red and orange colour) terms
up to $t^5$ are taken into account and the high-energy expansion
(light and dark blue) includes Pad\'e approximations with at least
$(m_t^2)^{49}$ and at most $(m_t^2)^{56}$ terms. In all cases quartic
terms in $m_H$ are included.  For this value of $p_T$ we observe
perfect agreement of the two expansions which can be seen in the black
and gray curves and the scale on the right side of the plot.  For
smaller values of $p_T$ the small-$t$ expansion is even more reliable,
and for larger values of $p_T$ the high-energy expansion.
This and similar plots for different values of $p_T$ (see
Ref.~\cite{Davies:2023vmj}) demonstrate that the combination of the
two expansions cover the whole phase space.


\section{First steps to three loops: Fermionic contribution for $t=0$}

We have seen in the previous Section that the $t$ expansion covers a
large part of the phase space. For this reason, in
Ref.~\cite{Davies:2023obx} the limits $t=0$ and $m_H=0$ has been
applied to three-loop $gg\to HH$ diagrams which contain a closed loop
with massless fermions. Even for the leading expansion term the
reduction problem is quite involved such that further improvements are
necessary in order to obtain the complete result. Subleading expansion
terms will be even significantly more involved.

\begin{figure}[t]
  \centering
  \includegraphics[width=0.9\textwidth]{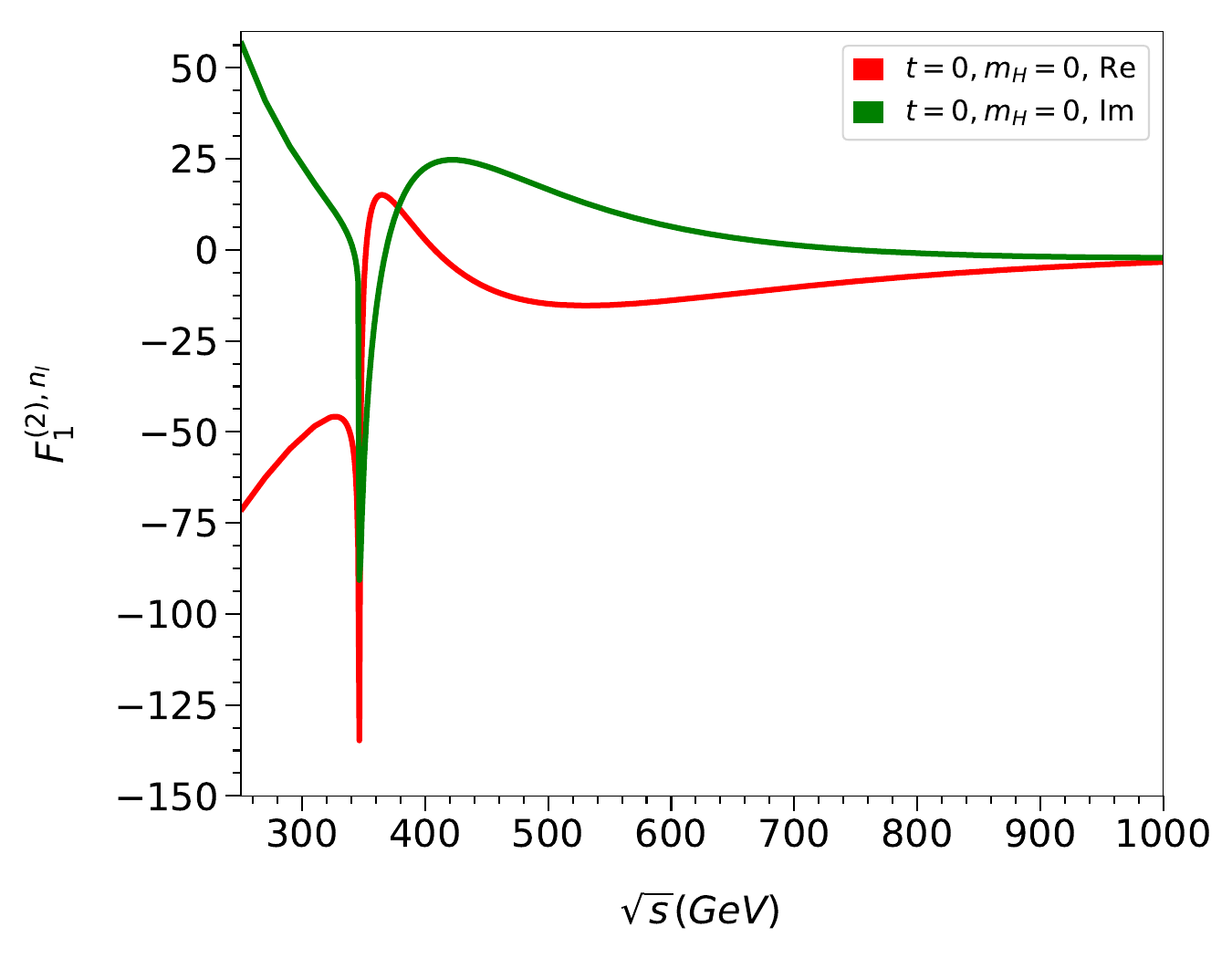}
  \caption{Real (red) and imaginary (green) parts of
    the $n_l$ part of $F_{\rm box1}$ at three loops.}
  \label{fig::F12_3nl}
\end{figure}

The first step taken in Ref.~\cite{Davies:2023obx} can be viewed as
exploratory study which shows that it is in principle possible to
apply the $t$ expansion at three loops. For $t=0$ and $m_H=0$ we have
performed the reduction~\cite{Klappert:2020nbg} to master integrals and
have established the corresponding differential equation.  Using
boundary conditions from the large-$m_t$ limit, which can be computed
analytically, we apply the "expand and match" approach to obtain
semi-analytic results, which are valid for all values of $s/m_t^2$. In
Fig.~\ref{fig::F12_3nl} we show results for the $n_l$ part of the form
factor $F_{\rm box1}$ (see Ref.~\cite{Davies:2023obx} for a precise
definition) as a function of $\sqrt{s}$. We stress that the plotted
expressions are stepwise defined functions where in each region we
have a power-log expansion with precise numerical coefficients.  The
choice for the expansion variable depends on the respective
region. For example, at threshold we have $v=\sqrt{1-4m_t^2/s}$ and at
high energies we have $x=m_t^2/s$.

Although $t=0$ and $m_H=0$ is a very crude approximation, from
considerations at one- and two-loop order
it can be expected that for $p_T$ in the vicinity of $100$~GeV
the (unknown) exact result is approximated 
at the level of 30\%. This is confirmed by a 
comparison at three-loop order in the large-$m_t$ limit,
where the complete $n_l$ terms are available~\cite{Davies:2019djw}.

At two loops the $t=0$ calculation provided the initial condition
for the  differential equations which could then be used to obtain
higher order expansion terms in $t$. At three loops this will
not be possible since the reduction to master integrals
of the box diagrams, which depend on $s,t$ and $m_t^2$, is currently
out of reach. Thus, we have to expand at the integrand level.


\section{Conclusions}

In this proceedings contribution we discuss several analytic
approximations for two- and three-loop corrections to $gg\to HH$,
which is used as a template for a wider range of processes like $gg\to
ZZ$, $gg\to ZH$ but also H plus jet production in gluon fusion.
The analytic results in the high-energy region and
the semi-analytic expression obtained for $t\to0$ are
sufficiently simple such that the fast numerical evaluation
is possible. Furthermore, in the combination
of both limits the whole phase space can be covered --- at least for
$gg\to HH$ at two loops.


\section*{Acknowledgements}  

I would like to thank Joshua Davies, Kay Sch\"onwald and Hantian Zhang for the
fruitful collaboration on the topics discussed in this contribution.  This
research was supported by the Deutsche Forschungsgemeinschaft (DFG, German
Research Foundation) under grant 396021762 --- TRR 257 ``Particle Physics
Phenomenology after the Higgs Discovery''.

\end{document}